\documentclass[aps,prl,twocolumn,showpacs,fleqn]{revtex4}

\usepackage{graphicx}
\usepackage{dcolumn}
\usepackage{bm}
\usepackage{psfrag}

\begin{document}

\title{Thickness-induced insufficient oxygen reduction
in La$_{2-x}$Ce$_{x}$CuO$_{4\pm\delta}$ thin films}

\author{B. X. Wu$^{1}$, K. Jin$^{1}$, J. Yuan$^{2}$, H. B. Wang$^{2}$,
T. Hatano$^{2}$, B. R. Zhao$^{1}$ and B. Y. Zhu$^{1}$}
\email{beiyi.zhu@aphy.iphy.ac.cn}

\affiliation{$^1$National Laboratory for Superconductivity,
Institute of Physics, and Beijing National Laboratory for Condensed
Matter Physics, Chinese Academy of Sciences, Beijing 100190, China}

\affiliation{$^2$National Institute for Materials Science, Tsukuba
305-0047, Japan}

\begin{abstract}
A series of electron-doped cuprate
La$_{2-x}$Ce$_{x}$CuO$_{4\pm\delta}$ thin films with different
thicknesses have been fabricated and their annealing time are
adjusted carefully to ensure the highest superconducting transition
temperature. The transport measurements indicate that, with the
increase of the film thickness ($< 100$~nm), the residual
resistivity increases and the Hall coefficient shifts in the
negative direction. Further more, the X-ray diffraction data reveal
that the c-axis lattice constant $c_0$ increases with the decrease
of film thickness. These abnormal phenomena can be attributed to the
insufficient oxygen reduction in the thin films. Considering the
lattice mismatching in the {\it ab}-plane between the SrTiO$_3$
substrates and the films, the compressive stress from the substrates
may be responsible for the more difficult reduction of the oxygen in
the thin films.
\end{abstract}


\maketitle

\pagebreak

\section{1. Introduction}
The electron-doped high-T$_c$ superconducting (SC) cuprates, e.g.
(Ln,Ce)$_2$CuO$_{4+\delta}$ (Ln=La, Pr, Nd, Sm), have been
extensively investigated since its discovery by Tokura {\it et al.}
in 1989~\cite{Discovery}. The hole-doped cuprates can exhibit
superconductivity when introducing a sufficient concentration of
hole carriers by either atom-substituting or oxygen doping in the
insulating Ln$_2$CuO$_4$ host material. However, the
superconductivity does not appear in the electron-doped cuprates
until a metastable T'-phase is achieved by an extra annealing
treatment of the as-grown samples. What happens during this
annealing process is still a controversy. Although oxygen reduction
is widely considered as the crucial factor in this annealing
process, it is hard to be determined stoichiometrically because the
amount of the oxygen loss $\delta$ is always relatively small. The
previous measurements of $\delta$ by either thermogravimetric
analysis (TGA)~\cite{Wang, Idemoto, lattice1, lattice2, Serquis} or
iodometric titration analyses~\cite{Suzuki, Singh, vlae} cannot come
to an agreement.

Many efforts have been made to clarify this puzzling issue. Based on
the transport and the thermopower measurements, Jiang {\it et al.}
proclaimed that both the redundant and the indigent of oxygen would
induce more impurities and enhance the scattering~\cite{Greene1,
Greene3, Greene4}. The lacking of oxygen can introduce a positive
contribution and the excess oxygen always results in a negative
contribution to the Hall coefficient, e.g. in
Nd$_{2-x}$Ce$_x$CuO$_4$ (NCCO) thin films and single crystals as
well as Pr$_{2-x}$Ce$_x$CuO$_4$ thin films~\cite{Greene5, Gauthier}.
Riou and Richard {\it et al.} studied the infrared transmission of
the Pr$^{3+}$ crystal-field in the Pr$_{2-x}$Ce$_x$CuO$_4$ single
crystal, and they pointed out that the oxygen in the CuO$_2$ plane
is partially removed during the reduction~\cite{Richard1,Richard2}
instead of the apical ones. Using X-ray and neutron-scattering
methods, Kang {\it et al.} have investigated the microscopic process
of the oxygen reduction in Pr$_{0.88}$LaCe$_{0.12}$CuO$_4$ single
crystals~\cite{Dai3}. They suggested that both the repair of Cu
deficiencies and the creation of oxygen vacancies can effectively
reduce the disorder and provide itinerant carriers for
superconductivity in the reduction treatment of the samples.
Yamamoto {\it et al.}~\cite{Naito1} claimed that if the reduction is
not sufficient in the NCCO thin films, the excess oxygen will occupy
the apical oxygen sites to compensate the Ce doping. While in the
excessive reduction films, the oxygen deficiencies appear at regular
oxygen sites in the CuO$_2$ plane. In our previous work, we have
also found that the extra annealing treatment can cause the Hall
coefficient shift in the positive direction in the dilute
cobalt-doped La$_{2-x}$Ce$_{x}$CuO$_{4}$ thin films~\cite{Kui2}.

Recently, the significant influence of the lattice constant on the
oxygen reduction has been investigated by varying the doping
concentration or substituting atoms with different
radii~\cite{lattice1, lattice2, Naito2, lattice3, lattice4}. It has
been found that the decrease of the lattice constant $a_0$ in the
{\it ab}-plane of the electron-doped cuprates can make the reduction
of excess oxygen more difficult.

In this paper, electron-doped La$_{2-x}$Ce$_{x}$CuO$_{4}$ (LCCO)
thin films with different thicknesses have been synthesized, and
their annealing time were adjusted carefully to ensure the highest
T$_c$ accordingly. Based on the transport measurement and the X-ray
diffraction analysis, we have found that the decrease of the
thickness may cause it difficult to create the oxygen vacancies in
the films during the annealing reduction.

\section{2. Experiments}
The optimally doped LCCO (x=0.105) thin films with a thickness $d$
varying from 17 to 600~nm were fabricated on the (100)-oriented
SrTiO$_3$ substrates by the dc magnetron sputtering
method~\cite{Zhao, WH, Kui4}. All the samples were annealed at
$550^\circ$C in vacuum lower than $10^{-3}$~Pa, and the annealing
time was adjusted to assure the sharp SC transition and the highest
T$_c$ according to their thicknesses. We have found that the optimal
annealing time is almost proportional to the film thickness $d$. For
the transport measurements, each sample was patterned into the
standard six-probe Hall bridge with 178 $\mu$m in length and 10
$\mu$m in width by photolithography and ion milling techniques. All
the measurements were carried out by Quantum Design PPMS-9
equipment. Table 1 shows the characteristics of a series of LCCO
thin films employed in the present work.

\begin{table}
\caption{\label{tabone}$d$, T$_{c0}$, T$_c^{onset}$ and $t$ for
various LCCO films. Here, $d$, T$_{c0}$, T$_c^{onset}$ and $t$ stand
for the thickness, the transition temperature of zero resistance,
the onset temperature of superconducting transition, and the optimal
annealing time for the sample with the designated thickness,
respectively.}
\begin{ruledtabular}
\begin{tabular}{@{}*{5}{l}}
Sample &$d$&T$_{c0}$&T$_c$$^{onset}$&$t$\\
No.&(nm)&(K)&(K)&(min)\\
\hline
1 & 17  & - & - & 12\\
2 & 33  & 21.5 & 26.5  & 20\\
3 & 67  & 26 & 27.5  & 23.1\\
4 & 100 & 26.5 & 28 & 26.8\\
5 & 150 & 26.5 & 28 & 35 \\
6 & 200  & 27.3 & 28.5 & 39 \\
7 & 600 & 27 & 28.5 & 70 \cr \\
\end{tabular}
\end{ruledtabular}
\end{table}

\section{3. Results and discussions}
\begin {figure}[!]  
\begin{center}
\includegraphics*[bb=54 0 433 375, width=8cm, clip]{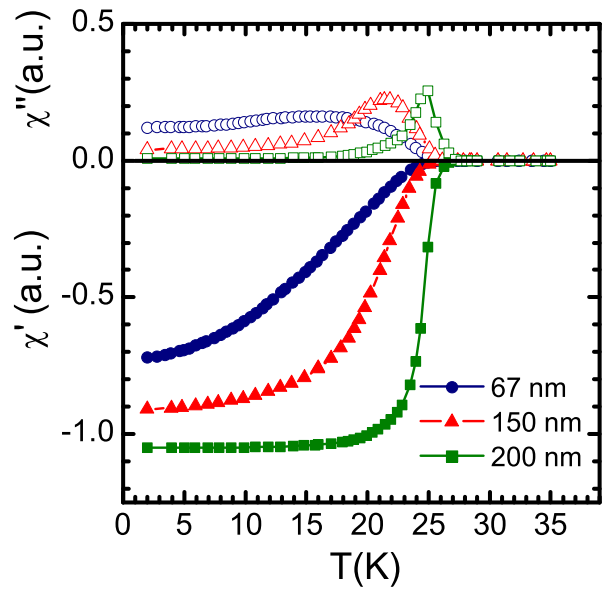}
\caption{\label{MT} Ac susceptibility $\chi'$ (solid dots) and
$\chi''$ (open circles) versus the temperature for the LCCO films
with different thicknesses. All the data of $\chi'$ and $\chi''$ for
different films have been normalized to their thicknesses.}
\end{center}
\end{figure}
%
Fig. \ref{MT} shows the ac susceptibility for the LCCO films with
different thicknesses. The ac susceptibility is measured in the
Quantum designed MPMS by the zero-field-cooling course and the ac
drive field is applied perpendicular to the {\it ab}-plane with 1~Oe
in amplitude and 333~Hz in frequency. Both the real and imaginary
components, i.e., $\chi'$ and $\chi''$, show that the film with
larger thickness has a sharper SC transition and a relatively high
T$_c$.
Additionally, the normalized shielding signal $\chi'$ increases with
the increase in thickness, which indicates that the SC component in
the thick films are larger than those in the thin ones.

\begin {figure}[!]  
\begin{center}
\includegraphics*[bb=0 0 468 358, width=8cm, clip]{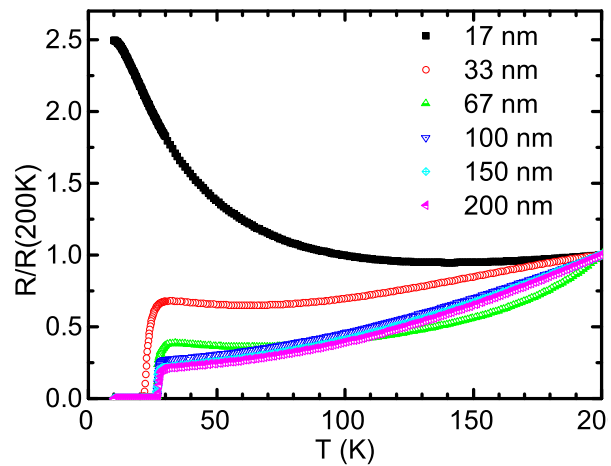}
\caption{\label{Tc}Temperature dependence of the resistance for LCCO
films with different thicknesses. All the data in each curve are
normalized to the resistance at 200~K respectively.}
\end{center}
\end{figure}
In order to investigate the effect of the film thickness on oxygen
reduction, we study the transport properties of these LCCO films.
Fig.~\ref{Tc} shows temperature dependence of the normalized
resistance for the samples with different thicknesses. When the film
thickness $d$ is larger than 100~nm, the curves are almost
overlapping. This indicates that all the samples have quite similar
temperature dependences. However, the ration of the residual
resistance (RRR), T$_{c0}$ and T$_c^{onset}$ decrease with the film
thickness decreasing from 100 nm to 33 nm. If the sample is
ultrathin, e.g. $d = 17$~nm, it becomes an insulator no matter how
we adjust the annealing time. The dependence of the resistance and
T$_c$ on the thickness can be attributed to the different influence
of the substrate on the films. The thinner the film is, the more
prominent the influence of the substrate on the properties is when
$d < 100$~nm. We propose that the influence of the substrate on the
properties of the films can be worked out by the oxygen reduction
during the annealing process. However, either the excess oxygen
caused by insufficient reduction or the oxygen vacancies induced by
the excessive reduction can act as impurities to enhance the
scattering and suppress the superconductivity~\cite{Greene3}. Since
Hall effect is an effective method to disclose the type of the
carriers, we try to measure the Hall coefficient to clarify the
controversy.

\begin {figure}[!]  
\begin{center}
\includegraphics*[bb=15 0 480 360, width=8cm, clip]{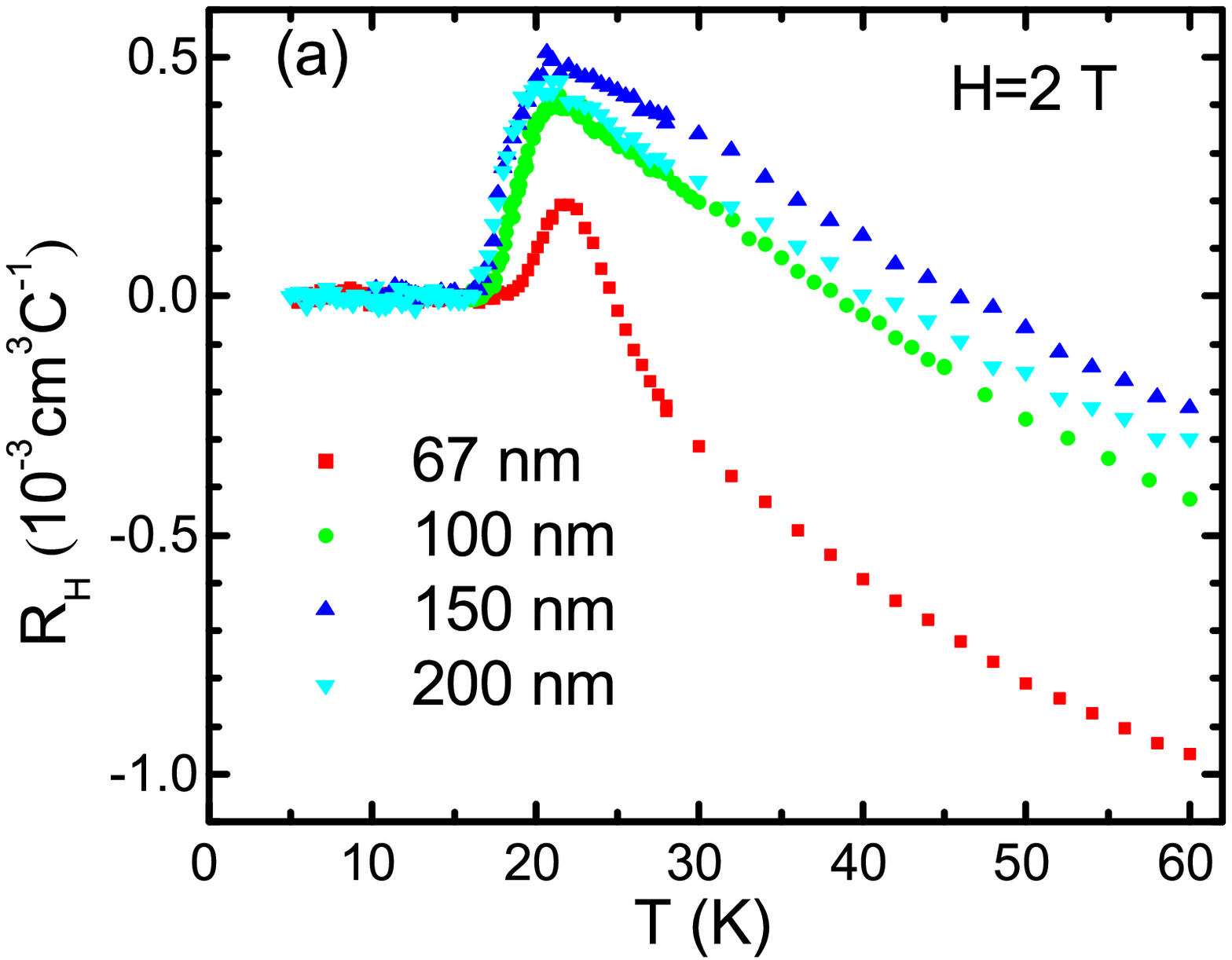}
\includegraphics*[bb=15 0 480 360, width=8cm, clip]{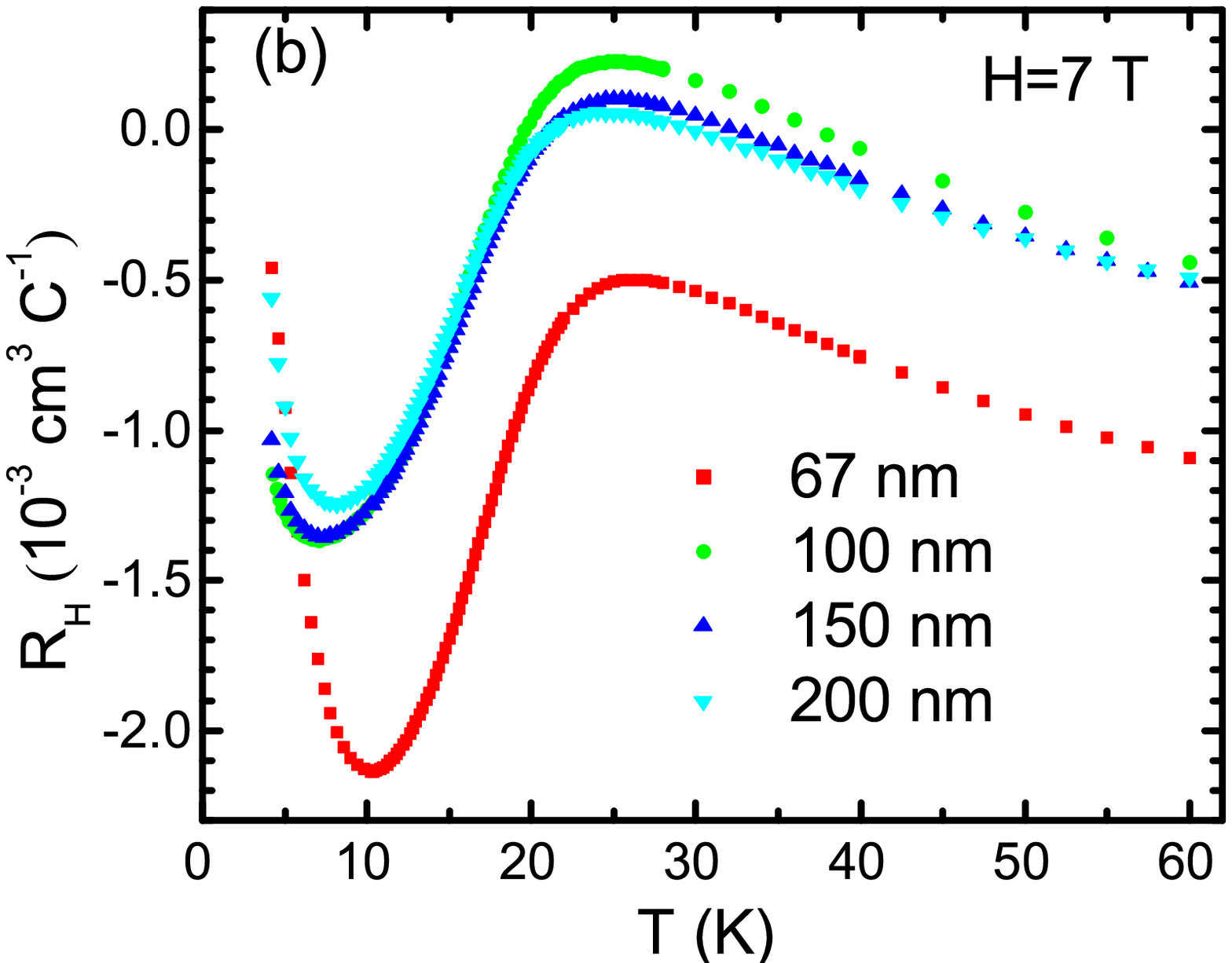}
\caption{\label{RH}The Hall coefficient $R_H$ versus the temperature
T for the LCCO thin films with different thicknesses at the magnetic
fields 2 and 7~T.}
\end{center}
\end{figure}
%
Fig.~\ref{RH} shows the Hall coefficient R$_H$ of the LCCO films
with different thickness. We find that all the samples have an
anomalous temperature dependence of R$_H$ with sign reversals. Since
there are two kinds of carriers in electron-doped
cuprates~\cite{Greene3, ARPES}, the Hall sign reversals versus the
magnetic fields and the temperatures can be attributed to
competition between the hole-like and the electron-like carriers
under the regime of the two-band model~\cite{Kui3, Hurd}. Here, we
will focus on the dependence of the Hall coefficient on the film
thickness. In Fig.~\ref{RH}(a) and (b) at H = 2~T and 7~T, the R$_H$
curves show quite a similar temperature dependence when $d \geq
100$~nm, respectively, while, in the case of $d = 67$~nm, the
nonzero Hall resistance R$_H$ shows clear bias from those of $d \geq
100$~nm and shifts to the negative. As mentioned
above~\cite{Greene3, Greene4, Greene5, Kui2}, the lack of oxygen
will introduce a positive hole-like contribution to the Hall
coefficient. Therefore, the enhancement of the negative R$_H$ of the
thin films with $d = 67$~nm can be attributed to the excess oxygen
due to the insufficient reduction process. Since the annealing time
of the films with different thickness are adjusted to their optimal
conditions to get the sharp transition and the highest T$_c$, we may
conclude that it is almost impossible to get rid of all the excess
oxygen in the thin films. The excess oxygen caused by insufficient
reduction in the thin films not only enhances the impurity
scattering but also results in the strong negative response of the
Hall coefficient.

\begin {figure}[!]  
\begin{center}
\includegraphics*[bb=20 150 460 668, width=8cm, clip]{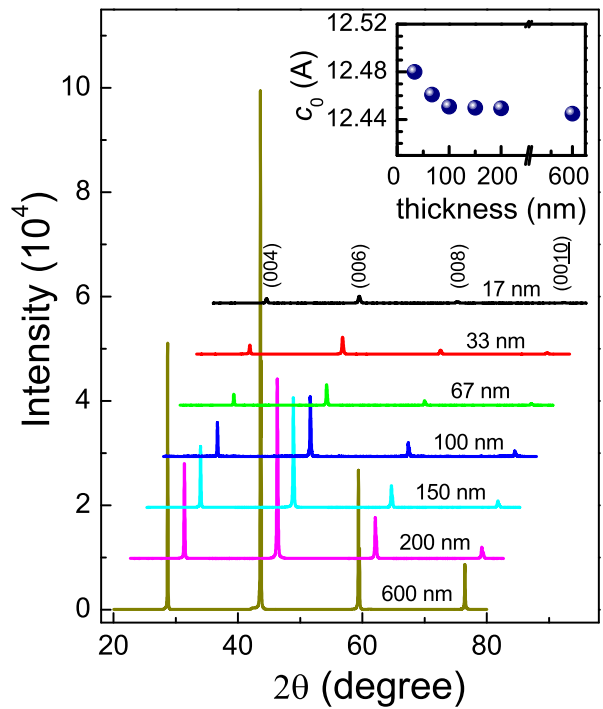}
\caption{\label{XRD} $\theta \sim 2 \theta$ X-ray diffraction
diagram for the LCCO films with different thicknesses. The substrate
peaks are removed from the curves. The inset shows the dependence of
the $c$-axis lattice length $c_{0}$ on the thickness.}
\end{center}
\end{figure}
It is important to make clear that the mismatch effect between the
substrate and the film originated from their different lattice
constants in the $ab$-plane, which may lead to the difficult oxygen
reduction in the ultrathin films as we have studied above.
Fig.~\ref{XRD} shows the $\theta \sim 2 \theta$ X-ray diffraction of
the LCCO films with the thickness varying from 17to 600~nm. All the
films exhibit good $c$-axis orientation and the amplitude of the
$(00l)$ peak increases monotonously with the film thickness. We
calculate the lattice constant $c_0$ along the $c$-axis and show its
dependence on the film thickness in the inset of Fig.~\ref{XRD}. We
find that the lattice constant $c_0$ does not depend on the
thickness when $d \geq 100$~nm, while if $d < 100$~nm, $c_0$
decreases with the increase in thickness, which can be understood as
that the apical oxygen is easy to escape from the LCCO samples with
the increase of the film thickness~\cite{Naito2}. We may notice that
the bond-length mismatch or the internal stress due to the small
rare-earth atom Ln$^{3+}$ substitution in the electron-doped
Ln$_{2-x}$Ce$_x$CuO$_4$ samples will cause it difficult for the
oxygen reduction as reported in Refs.~\cite{lattice2}
and~\cite{lattice3}. Since the lattice constants of $a_0$ are 3.905
and 4.010~\AA ~for the SrTiO$_3$ crystal and the optimal doped LCCO
film, respectively, the mismatch effect between them will be very
strong in the ultrathin LCCO films, e.g. $d < 100$~nm. Due to the
compression stress from the SrTiO$_3$ substrate, the lattice
constant $a_0$ of the ultrathin films probably decreases, which may
lead to it being more difficult for the oxygen reduction during the
annealing process.

Regarding the position the excess oxygen occupies in the lattice,
Higgins~\cite{Greene5} and Yamamoto~\cite{Naito1} agree on the
apical oxygen site, while Riou and Richard~\cite{Richard1, Richard2}
hold out for the O(1) site in the CuO$_2$ plane. Our previous
studies~\cite{Kui2} have indicated that the O(1) site vacancies can
be caused by excessive reduction in the annealing process for the
highest T$_c$. Here, we want to emphasize that the annealing process
on the thin films can remove the unwanted apical oxygen adequately,
but the excessive annealing would induce vacancies mostly in the
regular O(1) site.

\section {4. Conclusion}

We have investigated the transport properties including both the
longitudinal and the Hall resistances, as well as the X-ray
diffraction in the electron-doped LCCO thin films with different
thicknesses. With the decrease of the film thickness when $d <
100$~nm, the RRR decreases and the Hall coefficient shifts towards
the negative direction. The X-ray diffraction data reveal that the
c-axis lattice constant $c_0$ increases with the decrease of film
thickness. These nontrivial phenomena can be attributed to
insufficient oxygen reduction during the annealing course in the
ultrathin films. Due to the lattice mismatch effect between the
SrTiO$_3$ substrates and the thin films in the {\it ab}-plane, the
compressive stress from the substrates enhances the difficulty of
the oxygen reduction in the thin films when $d < 100$~nm.

\section{Acknowledgments}

We acknowledge the support from the MOST 973 project, the National
Nature Science Foundation and the SRF for ROCS, SEM Project of
China. B.X.W. is grateful to S. Arisawa for his help with the
transport measurement during her stay in NIMS.


\begin{thebibliography}{99}

\bibitem{Discovery}Tokura Y, Takagi T and Uchida S 1989 {\it Nature}
{\bf 337} 345

\bibitem{Wang}Wang E, Tarascon J -M, Greene L H, Hull G W and
McKinnon W R 1990 {\it Phys. Rev. B} {\bf 41} 6582

\bibitem{Idemoto}Idemoto Y and Fueki K 1991 {\it Jpn. J. Appl. Phys.}
{\bf 30} 2471

\bibitem{lattice1}Kawashima T and Takayama-Muromachi E 1994 {\it Physsica C}
{\bf 219} 389

\bibitem{lattice2}Zhu Y T and Manthiram A 1994 {\it Physica C}
{\bf 224} 256
\bibitem{Serquis}Serquis A, Prado F and Caneiro A 1999 {\it Physica C}
{\bf 313} 271

\bibitem{Suzuki}Suzuki K, Kishio K, Hasegawa T and Kitazawa K 1990
{\it Physica C} {\bf 166} 357

\bibitem{Singh}Singh O G, Padalia B D, Prakash Om, Suba K, Narlikar
A V and Gupta L C 1994 {\it Physica C} {\bf 219} 156

\bibitem{vlae}Vlaeminck H, Goossens H H, Mouton R, Hoste S and
Van Der Kelen G 1991 {\it J. Mater. Chem.} {\bf 1} 863

\bibitem{Greene1}Jiang W, Peng J L, Li Z Y and Greene R L 1993
{\it Phys. Rev. B} {\bf 47} 8151

\bibitem{Greene3}Jiang W, Mao S N, Xi X X, Jiang X, Peng J L,
Venkatesan T, Lobb C J and Greene R L 1994 {\it Phys. Rev. Lett.}
{\bf 73} 1291

\bibitem{Greene4}Xu X Q, Mao S N, Jiang W, Peng J L and Greene R L 1996
{\it Phys. Rev. B} {\bf 53} 871

\bibitem{Greene5}Higgins J S, Dagan Y, Barr M C, Weaver B D and
Greene R L 2006 {\it Phys. Rev. B} {\bf 73} 104510

\bibitem{Gauthier}Gauthier J, Gagn\'{e} S, Renaud J, Gosselin M -\`{E},
Fournier P and Richard P 2007 {\it Phys. Rev. B} {\bf 75} 024424

\bibitem{Richard1}Riou G, Richard P, Jandl S, Poirier M,
Fournier P, Nekvasil V, Barilo S N and Kurnevich L A 2004 {\it Phys.
Rev. B} {\bf 69} 024511

\bibitem{Richard2}Richard P, Riou G, Hetel I, Jandl S,
Poirier M and Fournier P 2004 {\it Phys. Rev. B} {\bf 70} 064513

\bibitem{Dai3}Kang H J, Dai P C, Campbell B J, Chupas P J,
Rosenkranz S, Lee P L, Huang Q Z, Li S L, Komiya S and Ando Y 2007
{\it Nat. Mater.} {\bf 6} 224

\bibitem{Naito1}Yamamoto H, Naito M and Sato H 1997 {\it Phys. Rev. B} {\bf
56} 2852

\bibitem{Kui2}Jin K, Yuan J, Zhao L, Wu H, Qi X Y, Zhu B Y, Cao L X, Qiu X G, Xu B,
Duan X F and Zhao B R 2006 {\it Phys. Rev.B} {\bf 74} 094518

\bibitem{Naito2}Matsumoto O, Utsuki A, Tsukada A, Yamamoto H, Manabe T
and Naito M 2008 {\it Int. Symposium on Superconductivity}

\bibitem{lattice3}Zhu Y T and Manthiram A 1994 {\it Phys. Rev. B} {\bf 49} 6293

\bibitem{lattice4}Fujita K, Noda T, Kojima K M, Eisaki H and
Uchida S 2005 {\it Phys. Rev. Lett.} {\bf 95} 097006

\bibitem{Zhao}Zhao L, Wu H, Miao J, Yang H, Zhang F C, Qiu X G and Zhao B R
2004 {\it Supercond. Sci. Technol.} {\bf 17} 1361

\bibitem{WH}Wu H, Zhao L, Yuan J, Cao L X, Zhong J P, Gao L J, Xu B, Dai P C, Zhu B Y,
Qiu X G and Zhao B R 2006 {\it Phys. Rev. B} {\bf 73} 104512

\bibitem{Kui4}Jin K, Zhu B Y, Wu B X, Gao L J and
Zhao B R 2008 {\it Phys. Rev. B} {\bf 78} 174521

\bibitem{ARPES}Armitage N P, et al., 2002 {\it Phys. Rev. Lett.} {\bf 88} 257001

\bibitem{Kui3}Jin K, Zhu B Y, Yuan J, Wu H, Zhao L, Wu B X, Han Y, Xu B, Cao L X,
Qiu X G and Zhao B R 2007 {\it Phys. Rev. B} {\bf 75} 214501

\bibitem{Hurd}Hurd C M 1972 {\it The Hall Effect in Metals and Alloys} (New
York: Plenum)

\end{thebibliography}
\end{document}